\documentclass[11pt,twoside]{article}
\usepackage{asp2010}

\resetcounters

\begin{document}

\title{Convective Core Overshoot \& Mass Loss in Classical Cepheids: A Solution to the Mass Discrepancy?}
\author{Hilding~R.~Neilson, Matteo~Cantiello, and Norbert~Langer
\affil{Argelander Institut f\"{u}r Astronomie, Universit\"{a}t Bonn}}

\begin{abstract}
We explore the role of mass loss and convective core overshoot in the evolution of Classical Cepheids.  Stellar evolution models are computed with a recipe for pulsation-driven mass loss and it is found that mass loss alone is unable to account for the long-standing Cepheid mass discrepancy.  However, the combination of mass loss and moderate convective core overshooting does provide a solution, bringing the amount of convective core overshooting in Cepheids closer to that found for other stars.
\end{abstract}

\section{Introduction}
Classical Cepheids are powerful laboratories for understanding stellar astrophysics, yet an important unanswered question is what are the masses of Cepheids.  Cepheid masses are determined using multiple different methods, including stellar evolution models matching measured effective temperatures and luminosities, stellar pulsation models matching pulsation periods and amplitudes, and from observations of binaries where one component is a Cepheid.  However, these three methods do not agree such that masses based on stellar pulsation models tend to be smaller than masses based on stellar evolution models.  Furthermore, dynamic masses are consistent with stellar pulsation masses \citep[e.g.][]{Evans2008}.  

This mass difference, called the Cepheid Mass Discrepancy, is a long-standing problem \citep{Cox1980} that is important not just for understanding Cepheids themselves but how stars evolve in general.  Because dynamic masses agree with the pulsation masses but not the stellar evolution masses then there must be physics missing from the stellar evolution models leading to Cepheid masses being overestimated  \citep{Keller2006,Keller2008}.  By finding the underlying source of the mass discrepancy, we can constrain the evolution of massive stars in general, leading to a better understanding of their later phases, such as asymptotic giant branch stars, white dwarfs and supernovae.

Historically, the Cepheid Mass Discrepancy was found to be about $40\%$ in general and about a factor of two for beat Cepheids \citep{Cox1980}.  However, \cite{Moskalik1992} reduced the mass discrepancy substantially when the OPAL opacities were included in stellar evolution models.  This result was a tremendous step forward in resolving the mass discrepancy. However, \cite{Caputo2005} and \cite{Keller2008} found that the mass discrepancy for Galactic Cepheids is still present, with typical values of $10$ - $20\%$.  Furthermore, \cite{Keller2006} modeled stellar pulsation of Galactic, and Large and Small Magellenic Cloud Cepheids and found that the mass discrepancy is a function of metallicity, increasing with decreasing metallicity.
The purpose of this work is to explore some of the physics used in stellar evolution calculations and test the impact of different input physics on the Cepheid mass discrepancy.  

\section{Possible Causes of the Mass Discrepancy}
There have been a number of possible solutions suggested to resolve the mass discrepancy \citep[see][for more details]{Bono2006}.  The four most probable causes are: missing opacity, rotational mixing, convective core overshooting (CCO), and mass loss.

\cite{Bono2006} argued that for opacity
changes to account for the discrepancy, they would
need to differ by a factor of two.  The difference between the Opacity Project \citep{Badnell2005} and previous opacities is at most $10\%$, suggesting that an opacity revision of a factor of two appears unlikely.  Therefore new opacities would only have a small impact on the mass discrepancy, and thus missing opacity is not a plausible solution.

Rotational mixing is a second possible solution.  A star that rotates rapidly during its main sequence evolution will mix hydrogen into the core. Thus its post main sequence helium core is more massive than if the same star evolved with negligible rotation.  The more massive helium core  would cause the star to be more luminous when the model crosses the Cepheid Instability Strip, 
and thus would predict a smaller mass for a Cepheid model with the same luminosity as a Cepheid model that does not include rotational mixing.  Furthermore, rotational mixing will change the surface abundances of Cepheids as well as the structure of the post main sequence blue loop evolution where stars undergo their second and third crossings of the Instability Strip \citep{Maeder2001}.  Rotational mixing is a possible solution to the mass discrepancy that should be explored further, however, it is beyond the scope of this work. 

\citet[][and references therein]{Keller2008} suggested convective core overshoot  (CCO) as yet another possible solution.   Convective core overshooting during the main-sequence evolution mixes hydrogen into the nuclear-burning region.  Just like rotational mixing, this will lead to a more massive helium core and a more luminous Cepheid.  The amount of CCO in stellar evolution models is often determined from the parameterization $\Lambda_c = \alpha_c H_P$, where $H_P$ is the pressure scale height, $\alpha_c$ is a free parameter, and $\Lambda_c$ is the distance above the convective region that overshooting penetrates.  \cite{Keller2006} and \cite{Keller2008} found that to solve the mass discrepancy $\alpha_c$ must equal $0.5$ - $1$. However, this range of $\alpha_c$ is larger than that found for other stars.

Mass Loss occurs throughout the entire evolution of a star and hence suggests another possible solution to the mass discrepancy.  However, mass loss during the main sequence and red giant stage of evolution of intermediate mass stars appears to negligibly contribute to the mass discrepancy \citep{Lanz1992, Willson2000} with the exception of the Be stars. \cite{Neilson2008, Neilson2009} developed a prescription for pulsation-enhanced mass loss during the Cepheid stage of evolution and found that the predicted Cepheid mass-loss rates are large enough to contribute to the mass discrepancy.  However, it is uncertain if mass loss completely solves the mass discrepancy. Further evidence of Cepheid mass loss includes observations of asymmetry of $H\alpha$ line profiles of Galactic Cepheids \citep{Nardetto2008} as well as from modelling the infrared excess of Large Magellanic Cloud Cepheids \citep{Neilson2010}. However, \cite{Keller2008} noted that mass loss needs to be as efficient for low-mass Cepheids as for large-mass Cepheids, which would be seem unlikely.

\section{Methodogy}
We test if mass loss during the Cepheid stage of stellar evolution is a plausible solution to the Cepheid mass discrepancy, either on its own or when coupled with various values for convective core overshoot, $\alpha_c$. We compute grids of stellar evolution models using a one-dimensional hydrodynamic stellar evolution code \citep{Heger2000} for stars with initial mass $M = 4, 6, 8,$ and $9~M_\odot$. These models are evolved with four different sets of input parameters:
\begin{enumerate}
\item no pulsation-driven mass loss and no CCO,
\item pulsation-driven mass loss, using the prescription from \cite{Neilson2008} and no CCO,
\item pulsation-driven mass loss and CCO, with $\alpha_c = 0.1$, and
\item pulsation-driven mass loss and CCO, with $\alpha_c = 0.335$, the value found by Brott et al. (in prep) \citep[see][for details]{Vink2010}.
\end{enumerate}
We compute the contribution of mass loss to the mass discrepancy $(\Delta M/M)$ by subtracting the initial mass of the stellar model from the mass at the end of Cepheid evolution, and divide by the initial mass. We calculate the contribution of CCO towards the mass discrepancy to be that $\alpha_c = 0.1$ leads to $\Delta M/M = 2.5\%$, $\alpha_c = 0.2$ is $\Delta M/M = 5\%$ and $\alpha_c = 0.335$ is $\Delta M/M = 8.375\%$ because of the change of luminosity due to increasing the amount of CCO.  The pulsation mass is computed from the luminosity, period, and other pulsation quantities; therefore we need to include the contribution of CCO in our computed mass difference. We adopt the mass of the stellar model at the end of Cepheid evolution as equivalent to the pulsation mass that was found by \cite{Keller2008} and the initial stellar mass with zero CCO is equivalent to the evolution mass based on the definition used by \cite{Caputo2005} and \cite{Keller2008}.  Therefore, this measurement is a prediction of the maximum contribution of the chosen input physics on the Cepheid mass discrepancy.  Note that we assume pulsation-driven mass loss occurs only on the Cepheid Instability Strip.   

\section{Results}
\begin{figure}[t]
\begin{center}
\includegraphics[width = 0.5\textwidth]{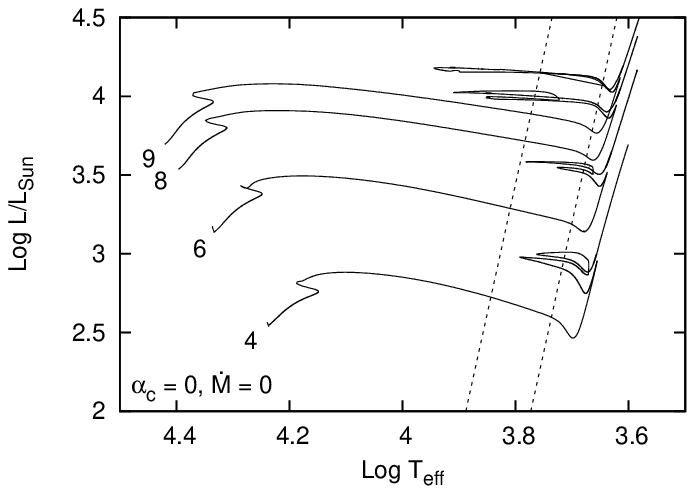}\includegraphics[width = 0.5\textwidth]{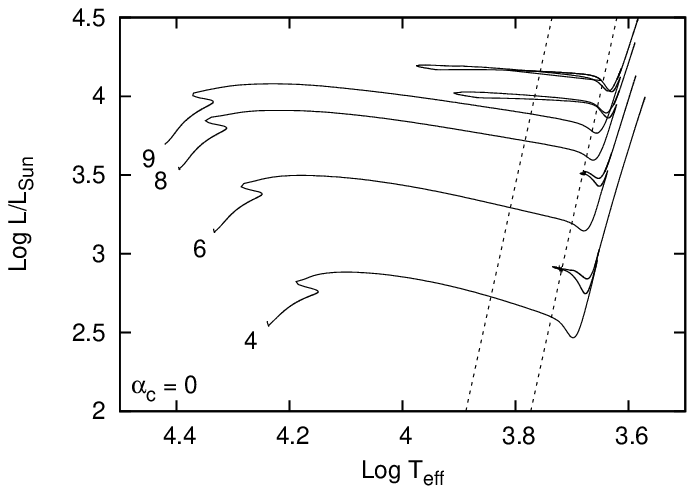}
\includegraphics[width = 0.5\textwidth]{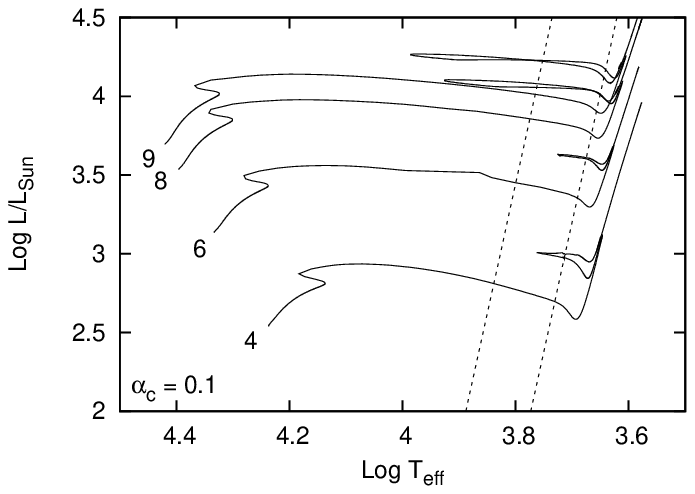}\includegraphics[width = 0.5\textwidth]{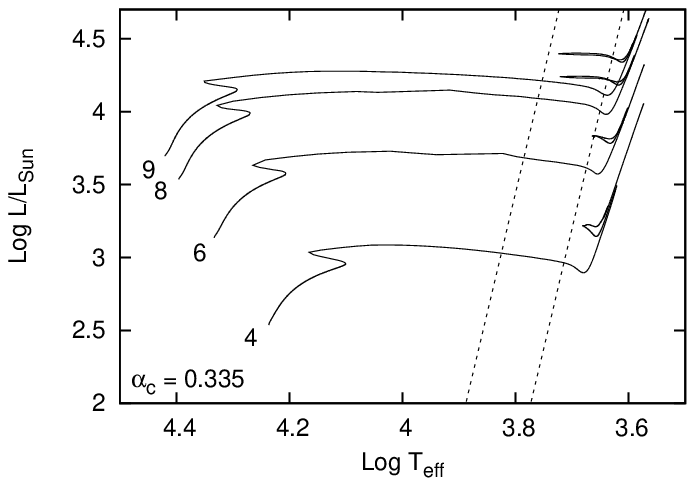}
\end{center}
\caption{Stellar evolutionary tracks for models with initial mass $M = 4,6,8,9~M_\odot$ and for the four scenarios discussed in the text. (Upper Left) no pulsation-driven mass loss nor CCO, (Upper Right) with pulsation-driven mass loss and zero CCO, (Lower Left) with pulsation-driven mass loss and CCO, $\alpha_c = 0.1$, and (Lower Right) with pulsation-driven mass loss and CCO, $\alpha_c = 0.335$. The dashed lines are the borders of the Cepheid Instability Strip.}
\label{f1}
\end{figure}

The stellar evolution tracks for each case and mass are shown in Figure \ref{f1} and the computed contribution to the mass discrepancy is shown in Table \ref{t1}. For the $\alpha_c = 0.335$ case, the $6,8$, and $9~M_\odot$ stars lose enough mass during the Cepheid stage of evolution to be consistent with the mass discrepancy determined by \cite{Keller2008}.  Every stellar model undergoes blue loop evolution, where the width of the blue loop is determined by effective temperature range spanned, increases with initial mass implying that the stars undergo the Cepheid stage of evolution for each case.  We do not model the radial pulsations but assume that the star is undergoing Cepheid evolution if the effective temperature and luminosity of the stellar model is inside the Cepheid Instability Strip.

Once a stellar model enters the Instability Strip, we use the \cite{Neilson2008} prescription for mass loss where radial pulsation in the star generates shocks that enhance the wind.  In this model, the mass-loss rate is determined by the stellar mass, luminosity, effective temperature, and pulsation period, where the period is computed from a Period-Mass-Radius relation \citep{Gieren1989}. This relation has a predicted period error of approximately $25\%$ which may lead to a mass-loss rate uncertainty of approximately a few hundred percent. \cite{Neilson2008} found that mass-loss rates of Galactic Cepheids ranges from $10^{-11} - 10^{-7}~M_\odot yr^{-1}$ and that mass-loss rates are potentially increasing with lower metallicity \citep{Neilson2009}. We apply this prescription to stellar evolution models and explore how mass loss affects blue loop evolution.
\begin{table}[t]
\caption{The contribution of mass loss and CCO to the Cepheid Mass Discrepancy.}\label{t1}
\begin{center}
\begin{tabular}{|c|c|c|c|c|}
\hline
Initial Mass $(M_\odot)$ & \hspace{0.2cm}Case 1 \hspace{0.2cm} & \hspace{0.2cm}Case 2\hspace{0.2cm} & \hspace{0.2cm}Case 3 \hspace{0.2cm}& \hspace{0.2cm}Case 4\hspace{0.2cm} \\
\hline
$4$ & $0\%$&$6.25\%$&$6.00\%$ & $ 8.72\%$ \\
$6$ & $0\%$&$3.00\%$&$8.30\%$ & $ 18.54\%$ \\
$8$ & $0\%$&$1.75\%$&$4.50\%$ & $ 16.24\%$ \\
$9$ & $0\%$&$1.67\%$&$4.40\%$ & $ 15.00\%$\\
\hline
\end{tabular}
\end{center}
\end{table}

\begin{figure}[t]
\begin{center}
\includegraphics[width = 0.95\textwidth]{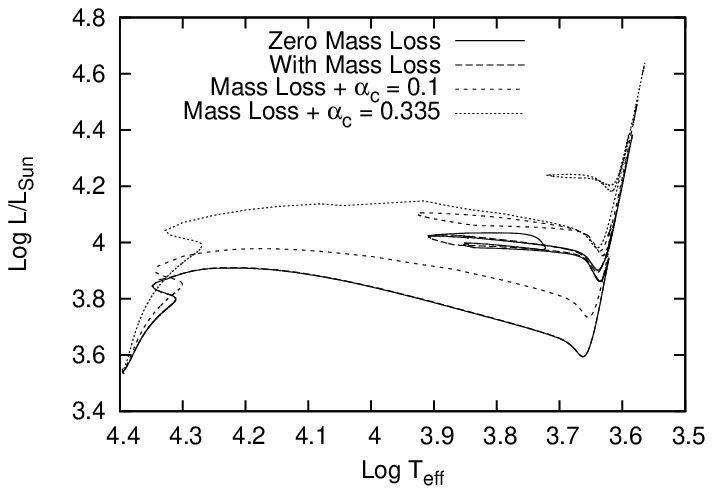}
\end{center}
\caption{Comparison of the $8~M_\odot$ stellar evolution tracks for the four scenarios explored in this work. The location and structure of the blue loops change due to the different input parameters.}
\label{f3}
\end{figure}

Consider first the $8~M_\odot$ stellar evolution models that are shown in Figure \ref{f3}. The tracks for models with and without Cepheid mass loss and zero CCO have blue loops spanning identical ranges of effective temperature and luminosity suggesting that mass loss alone has little affect on the evolution of the star. When CCO is considered then the location of the blue loop shifts to larger luminosities. For example, the blue loop of the stellar evolution model with  $\alpha_c = 0.1$ has a luminosity that is about $0.1$ dex larger than the models without CCO while the model with $\alpha_c = 0.335$ is about $0.25$ dex more luminous than the models without CCO.

However, it is not clear how mass loss changes with different amounts of CCO and thus how mass loss affects structure of the blue loop. For instance, the mass-loss rate depends on the ratio of luminosity and mass, $L/M$, which increases with $\alpha_c$. On the other hand, the predicted mass-loss rate increases with decreasing pulsation period which in turn increases with stellar radius.  The radius increases with luminosity and thus with increasing $\alpha_c$.  Thus a change of CCO leads to changes in fundamental parameters that work against each other in determining the mass-loss rate.  In Figure \ref{f3}, the only instance where mass loss appears to affect the evolution of the star is for the case where $\alpha_c = 0.335$. 

\begin{figure}[t]
\begin{center}
\includegraphics[width =0.5\textwidth]{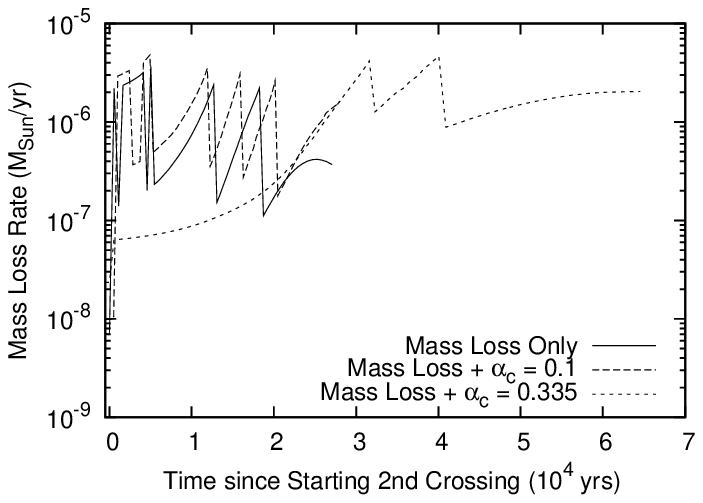}\includegraphics[width =0.5\textwidth]{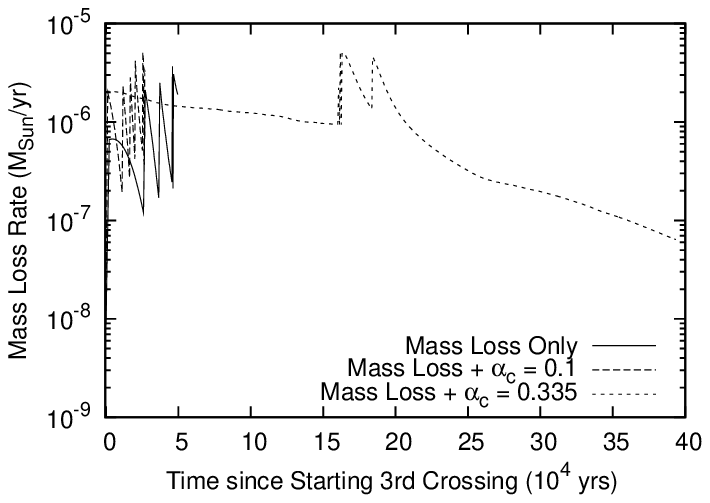}
\end{center}
\caption{The pulsation-driven mass-loss rates predicted for the $8~M_\odot$ models for the second and third crossings of the Cepheid Instability Strip and for the three scenarios that include pulsation-driven mass loss. The pulsation-driven mass-loss rate for the $8~M_\odot$ during its first crossing is not shown because the change of mass is too small to contribute significantly.}\label{f2}
\end{figure}

We explore this further in Figure \ref{f2}, where we compare the predicted mass-loss rates for the $8~M_\odot$ stellar evolution models for three cases where pulsation-driven mass loss is assumed for the second and third crossings of the Instability Strip; the first crossing is too short to contribute a significant mass change in the $8~\odot$ models.  During the second crossing, the timescale is about twice as long for $\alpha_c = 0.335$ as opposed to $\alpha_c = 0$ and $0.1$, yet there are little difference in mass-loss rates.  However, during the third crossing,  the $\alpha_c = 0.335$ model has a crossing time that is almost an order-of-magnitude longer than the cases with less CCO and has an average mass-loss rate
that is a factor of 2-3 larger than for the other cases.  

The predicted average mass-loss rate tends to increase with $\alpha_c$ when the star is evolving along the Instability Strip. For increasing CCO, then the ratio of the luminosity and mass, $L/M$, increases for Cepheids. Hence the amount of radiation pressure acting on a wind is larger and plays a bigger role in driving the wind than the change of pulsation period.  This is consistent with the results of \cite{Neilson2009}.
The change in timescale that the $8~M_\odot$ stars spend on the Cepheid Instability Strip is also not directly due to the amount of CCO but on the mass-loss rate and its dependence on CCO.  We hypothesize that the width of the blue loop decreases if the mass-loss rate is large, similar to the result found by \cite{Willson1986}, thus ``trapping'' the star in the Cepheid Instability Strip.  During the second crossing, a Cepheid's effective temperature is increasing and its radius is decreasing with time.  However, mass loss acts against the contraction of the stellar envelope, and ``puffs up'' the envelope.  The more inflated envelope prevents the effective temperature from increasing at the same rate as the cases with smaller CCO and mass loss. 

We find that the $9~M_\odot$ models appear to behave similarly as the $8~M_\odot$ models.  The width of the blue loop of the $9~M_\odot$ models is smaller for the $\alpha_c = 0.335$ case and is also trapped in the Instability Strip.  The stellar evolution tracks appear unchanged for smaller values of $\alpha_c$.  The $4$ and $6~M_\odot$ stellar models behave differently.  The models with $\alpha_c = 0$ and zero pulsation-driven mass loss appear to undergo multiple blue loops and when pulsation mass loss is included the multiple loops disappear and the one blue loop has a smaller width for both models. The $4$ and $6~M_\odot$ blue loops are more sensitive to mass loss relative to the $8$ and $9~M_\odot$.

The $4$ and $6~M_\odot$ model blue loops are also sensitive to CCO.  The blue loops for the $\alpha_c = 0.1$ models span a larger range of effective temperatures than the models with $\alpha_c = 0$.  However, the $\alpha_c = 0.335$ models have a smaller blue loop, so much so that the $4~M_\odot$ model blue loop does not enter the Cepheid Instability Strip.  This behavior is associated with the minimum mass at which stellar evolution models can form a blue loop.  The minimum mass  increases with increasing $\alpha_c$, such that at $\alpha_c = 0.335$ the $4~M_\odot$ model is approximately the minimum mass.  The $6~M_\odot$ model blue loop appears truncated at $\alpha_c = 0.335$ due to the combination of mass loss and convective core overshooting just like the $8$ and $9~M_\odot$ models.


\section{Summary}
The results of this work suggest that the Cepheid mass discrepancy can be resolved by the combination of pulsation-enhanced mass loss \citep{Neilson2008} on the Cepheid Instability Strip and convective core overshoot in the main sequence progenitors.  On its own, CCO is not a plausible solution because of measurements of the amount of CCO in other stars.  While this would require $\alpha_c > 0.5$ \citep{Keller2008}, from eclipsing binaries, the value of $\alpha_c = 0.2-0.4$ \citep{Clausen2010, Sandberg2010}, from $\beta$ Cephei stars $\alpha_c = 0.28 \pm 0.1$ \citep{Lovekin2010} while from early B-type stars $\alpha_c=0.335$ \citep{Hunter2008, Vink2010}.  Furthermore, large values of $\alpha_c$ may act to suppress the blue loop evolution in the most massive Cepheids. \cite{Kippenhahn1991} noted that whether stars go through blue loop evolution  depends on the mass and radius of the stellar core, the mass and change of hydrogen abundance of the hydrogen-burning shell.  A more massive stellar helium core, such as that produced when CCO is included, acts to quench blue loop evolution.

 Mass loss, on its own, also is not a plausible solution because the Cepheid crossing timescales are too short, and/or the mass-loss rates are too small.  However, the combination of CCO, with $\alpha_c=0.2$ - $0.4$, and pulsation-driven mass loss together provide a plausible solution.
   This work is preliminary and we intend to explore a larger mass range up to $M = 15 M_\odot$ as well as Small and Large Magellanic Cloud metallicities to test the metallicity dependence of the mass discrepancy found by \cite{Keller2006}. 
   
\acknowledgements
HRN is grateful for financial support from the Alexander von Humboldt Foundation. 
\bibliographystyle{asp2010}

\bibliography{neilson_h}

\begin{thebibliography}{}
\expandafter\ifx\csname natexlab\endcsname\relax\def\natexlab#1{#1}\fi
\expandafter\ifx\csname url\endcsname\relax
  \def\url#1{\texttt{#1}}\fi
\expandafter\ifx\csname urlprefix\endcsname\relax\def\urlprefix{URL }\fi
\providecommand{\eprint}[2][]{\url{#2}}

\bibitem[{{Badnell} et~al.(2005){Badnell}, {Bautista}, {Butler}, {Delahaye},
  {Mendoza}, {Palmeri}, {Zeippen}, \& {Seaton}}]{Badnell2005}
{Badnell}, N.~R., {Bautista}, M.~A., {Butler}, K., {Delahaye}, F., {Mendoza},
  C., {Palmeri}, P., {Zeippen}, C.~J., \& {Seaton}, M.~J. 2005, \mnras, 360,
  458

\bibitem[{{Bono} et~al.(2006){Bono}, {Caputo}, \& {Castellani}}]{Bono2006}
{Bono}, G., {Caputo}, F., \& {Castellani}, V. 2006, MmSAI, 77, 207

\bibitem[{{Brunish} \& {Willson}(1987)}]{Willson1986}
{Brunish}, W.~M., \& {Willson}, L.~A. 1987, in Stellar Pulsation, edited by
  {A.~N.~Cox, W.~M.~Sparks, \& S.~G.~Starrfield}, vol. 274 of Lecture Notes in
  Physics, Berlin Springer Verlag, 27

\bibitem[{{Caputo} et~al.(2005){Caputo}, {Bono}, {Fiorentino}, {Marconi}, \&
  {Musella}}]{Caputo2005}
{Caputo}, F., {Bono}, G., {Fiorentino}, G., {Marconi}, M., \& {Musella}, I.
  2005, \apj, 629, 1021. \eprint{arXiv:astro-ph/0505149}

\bibitem[{{Clausen} et~al.(2010){Clausen}, {Frandsen}, {Bruntt}, {Olsen},
  {Helt}, {Gregersen}, {Juncher}, \& {Krogstrup}}]{Clausen2010}
{Clausen}, J.~V., {Frandsen}, S., {Bruntt}, H., {Olsen}, E.~H., {Helt}, B.~E.,
  {Gregersen}, K., {Juncher}, D., \& {Krogstrup}, P. 2010, \aap, 516, A42+

\bibitem[{{Cox}(1980)}]{Cox1980}
{Cox}, A.~N. 1980, \araa, 18, 15

\bibitem[{{Evans} et~al.(2008){Evans}, {Schaefer}, {Bond}, {Bono}, {Karovska},
  {Nelan}, {Sasselov}, \& {Mason}}]{Evans2008}
{Evans}, N.~R., {Schaefer}, G.~H., {Bond}, H.~E., {Bono}, G., {Karovska}, M.,
  {Nelan}, E., {Sasselov}, D., \& {Mason}, B.~D. 2008, \aj, 136, 1137

\bibitem[{{Gieren}(1989)}]{Gieren1989}
{Gieren}, W.~P. 1989, \aap, 225, 381

\bibitem[{{Heger} et~al.(2000){Heger}, {Langer}, \& {Woosley}}]{Heger2000}
{Heger}, A., {Langer}, N., \& {Woosley}, S.~E. 2000, \apj, 528, 368

\bibitem[{{Hunter} et~al.(2008){Hunter}, {Brott}, {Lennon}, {Langer}, {Dufton},
  {Trundle}, {Smartt}, {de Koter}, {Evans}, \& {Ryans}}]{Hunter2008}
{Hunter}, I., {Brott}, I., {Lennon}, D.~J., {Langer}, N., {Dufton}, P.~L.,
  {Trundle}, C., {Smartt}, S.~J., {de Koter}, A., {Evans}, C.~J., \& {Ryans},
  R.~S.~I. 2008, \apjl, 676, L29. \eprint{0711.2267}

\bibitem[{{Keller}(2008)}]{Keller2008}
{Keller}, S.~C. 2008, \apj, 677, 483

\bibitem[{{Keller} \& {Wood}(2006)}]{Keller2006}
{Keller}, S.~C., \& {Wood}, P.~R. 2006, \apj, 642, 834

\bibitem[{{Kippenhahn} \& {Weigert}(1990)}]{Kippenhahn1991}
{Kippenhahn}, R., \& {Weigert}, A. 1990, {Stellar Structure and Evolution}

\bibitem[{{Lanz} \& {Catala}(1992)}]{Lanz1992}
{Lanz}, T., \& {Catala}, C. 1992, \aap, 257, 663

\bibitem[{{Lovekin} \& {Goupil}(2010)}]{Lovekin2010}
{Lovekin}, C.~C., \& {Goupil}, M. 2010, \aap, 515, A58+

\bibitem[{{Maeder} \& {Meynet}(2001)}]{Maeder2001}
{Maeder}, A., \& {Meynet}, G. 2001, \aap, 373, 555

\bibitem[{{Moskalik} et~al.(1992){Moskalik}, {Buchler}, \&
  {Marom}}]{Moskalik1992}
{Moskalik}, P., {Buchler}, J.~R., \& {Marom}, A. 1992, \apj, 385, 685

\bibitem[{{Nardetto} et~al.(2008){Nardetto}, {Groh}, {Kraus}, {Millour}, \&
  {Gillet}}]{Nardetto2008}
{Nardetto}, N., {Groh}, J.~H., {Kraus}, S., {Millour}, F., \& {Gillet}, D.
  2008, \aap, 489, 1263

\bibitem[{{Neilson} \& {Lester}(2008)}]{Neilson2008}
{Neilson}, H.~R., \& {Lester}, J.~B. 2008, \apj, 684, 569

\bibitem[{{Neilson} \& {Lester}(2009)}]{Neilson2009}
--- 2009, \apj, 690, 1829

\bibitem[{{Neilson} et~al.(2010){Neilson}, {Ngeow}, {Kanbur}, \&
  {Lester}}]{Neilson2010}
{Neilson}, H.~R., {Ngeow}, C., {Kanbur}, S.~M., \& {Lester}, J.~B. 2010, \apj,
  716, 1136

\bibitem[{{Sandberg Lacy} et~al.(2010){Sandberg Lacy}, {Torres}, {Claret},
  {Charbonneau}, {O'Donovan}, \& {Mandushev}}]{Sandberg2010}
{Sandberg Lacy}, C.~H., {Torres}, G., {Claret}, A., {Charbonneau}, D.,
  {O'Donovan}, F.~T., \& {Mandushev}, G. 2010, \aj, 139, 2347

\bibitem[{{Vink} et~al.(2010){Vink}, {Brott}, {Gr{\"a}fener}, {Langer}, {de
  Koter}, \& {Lennon}}]{Vink2010}
{Vink}, J.~S., {Brott}, I., {Gr{\"a}fener}, G., {Langer}, N., {de Koter}, A.,
  \& {Lennon}, D.~J. 2010, \aap, 512, L7+

\bibitem[{{Willson}(2000)}]{Willson2000}
{Willson}, L.~A. 2000, \araa, 38, 573

\end{thebibliography}

\end{document}